\documentclass[preprint,showpacs,preprintnumbers,amsmath,amssymb,endfloats*]{revtex4}
\usepackage{graphicx}

\begin{document}
\def \ee {\varepsilon}
\thispagestyle{empty}
\title{
Comment on ``Possible resolution of the Casimir force finite
temperature correction ``controversies''\,''
}

\author{R.~S.~Decca,${}^1$ E.~Fischbach,${}^2$
B.~Geyer,${}^3$ 
G.~L.~Klimchitskaya,${}^3$\footnote{on leave from
North-West Technical University, St.Petersburg, Russia}
D.~E.~Krause,${}^{4,2}$ D.~L\'{o}pez,${}^5$
U.~Mohideen,${}^6$
and V.~M.~Mostepanenko${}^3$\footnote{on leave from Noncommercial Partnership
``Scientific Instruments'',  Moscow,  Russia}
}

\affiliation{
${}^1$Department of Physics, Indiana University-Purdue
University Indianapolis, Indianapolis, Indiana 46202, USA\\
${}^2$Department of Physics, Purdue University, West Lafayette, Indiana
47907, USA\\
${}^3$Center of Theoretical Studies and Institute for Theoretical
Physics, Leipzig University,
D-04009, Leipzig, Germany \\
${}^4$Physics Department, Wabash College, Crawfordsville, Indiana 47933,
USA\\
${}^5$Bell Laboratories, Lucent Technologies, Murray Hill,
New Jersey 07974, USA \\
${}^6$Department of Physics and Astronomy, University of California,
Riverside, California 92521, USA
}

\begin{abstract}
The recently suggested  modification of the transverse electric 
contribution to the Lifshitz formula (S.\ K.\ Lamoreaux, 
arXiv:0801.1283) is discussed. We show that this modification 
is inconsistent with the data of two precise experiments, and 
violates the Nernst heat theorem. The preprint's suggestion 
concerning the  resolution of the ``apparent violation of the 
Third Law of Thermodynamics'' is shown to be incorrect. 
\end{abstract}

\maketitle

The preprint \cite{1} suggests a modification for the contribution
of the transverse electric modes to the Lifshitz free energy in the
configuration of two conducting semispaces at a temperature $T$ separated
by a gap of thickness $d$
\begin{equation}
{\cal F}^{\rm TE}=\frac{k_BT}{2\pi}\sum_{n=0}^{\infty}
{\vphantom{\sum}}^{\prime}\int_{0}^{\infty}\!\!\!qdq\ln\left[1-
\left(\frac{\gamma_{1n}-\gamma_{0n}}{\gamma_{1n}+\gamma_{0n}}\right)^2
{\rm e}^{-2\gamma_{0n}d}\right].
\label{eq1}
\end{equation}
\noindent
Here, $\gamma_{0n}^2=q^2+\xi_n^2/c^2$, $\xi_n$ are the Matsubara
frequencies, and the standard
expression $\gamma_{1n}^2=q^2+\varepsilon({\rm i}\xi_n)\xi_n^2/c^2$
\cite{2} is replaced with
\begin{equation}
\gamma_{1n}^2=q^2+\lambda^{-2}+\varepsilon({\rm i}\xi_n)\frac{\xi_n^2}{c^2},
\quad
\lambda^{-2}=\frac{e^2c_t}{\bar{\ee}\epsilon_0k_BT},
\label{eq2}
\end{equation}
\noindent
where $\lambda$ is the Debye-H\"{u}ckel screening length [$\ee(\omega)$
is the permittivity of a conductor, $c_t$ is the total carrier concentration,
$\epsilon_0$ is the permittivity of vacuum, 
and $\bar{\ee}$ is the dielectric
constant due to core electrons]. According to Ref.~\cite{1}, the effect of 
Debye-H\"{u}ckel screening in accordance with Eq.~(\ref{eq2}) leads to the 
same zero-frequency contribution in Eq.~(\ref{eq1}) as for ideal metals and
resolves the  contradiction between the Lifshitz formula combined with
the Drude model and the experimental data of 
Ref.~\cite{3}. (Note that there was a discussion \cite{3a,3b}
about the comparison between the experimental data of Ref.~\cite{3} 
and theory.)
Below we demonstrate, however, that the suggested modification is not only
inconsistent with two other more precise experiments on the measurement of the
Casimir force but also violates the Nernst heat theorem.

Using the approach of Ref.~\cite{1}, we have computed the Casimir pressure
between two Au plates and the Casimir force between an Au sphere and a
Si plate in the experimental configurations of Refs.~\cite{4,5}.
The values $c_t\approx 5.9\times 10^{22}\,\mbox{cm}^{-3}$ and
$c_t\approx 3.2\times 10^{20}\,\mbox{cm}^{-3}$ at $T=300\,$K
were used in the computations for
Au \cite{4} and highly doped $n$-type Si \cite{5}, respectively. 
 We have included the
term $\lambda^{-2}$, as in Eq.~(\ref{eq2}), only for $n=0$ (as noted in
\cite{1}, $\lambda$ is frequency-independent only at low frequencies
$<10^{10}\,$Hz). The inclusion of any nonzero $\lambda^{-2}$
in the Matsubara terms with $n\geq 1$ would only increase the magnitudes of 
the theoretical pressure and force, and thus would
increase the disagreement  between
experiment and theory. The standard
contribution of the transverse magnetic modes was employed \cite{2}. 
In Fig.\ 1(a) the differences between 
the computed theoretical Casimir pressures
and the experimental data of Ref.~\cite{4} are shown as dots at different
separations. In Fig.\ 1(b) the differences between the computed theoretical 
Casimir forces and respective data of Ref.~\cite{5} are presented. Solid
lines in both figures indicate the boundaries of the 95\% confidence intervals.
Dashed lines in Fig.\ 1(b) show the boundaries of 
the 70\% confidence intervals. 
As it is seen in Fig.\ 1, the experiment \cite{4} excludes the theoretical 
approach of Ref.~\cite{1} within the
separation region from 170 to 450\,nm at a 95\% confidence level. 
The experiment \cite{5} excludes this approach with a 95\% confidence at
separations from 62 to 82\,nm and with a 70\% confidence within a
wider separation region from 62 to 100\,nm.

Reference \cite{1} claims to solve 
the thermodynamic inconsistency in the
theory of the thermal Casimir force. As proved in Ref.~\cite{6}, and 
independently confirmed in \cite{7}, the Lifshitz formula combined with
the Drude model violates the Nernst heat theorem in the case of perfect
crystal lattices. The statement of Ref.~\cite{1} to the contrary is based 
on errors. Reference ~\cite{1} does not take into account the fact
that $T$ appears in the Casimir force calculation not only
through the factor $\exp(\hbar\omega/k_BT)$ but also through the
temperature-dependent dielectric permittivity $\ee(\omega,T)$.
We emphasize that this is actually the case for the Drude model. The behavior
of the free energy when $T$ goes to zero can be investigated
using the Abel-Plana formula
\begin{equation}
\sum_{n=0}^{\infty}{\vphantom{\sum}}^{\prime}f(n)=\int_{0}^{\infty}f(t)dt
+{\rm i}\int_{0}^{\infty}\frac{f({\rm i}t)-f(-{\rm i}t)}{{\rm e}^{2\pi t}-1}dt.
\label{eq3}
\end{equation}
\noindent
The second term on the right-hand side of this equation is not taken into 
account in \cite{1}. 
Although it goes to zero when $T$ vanishes, its
derivative  with respect to $T$ may not vanish when $T$
goes to zero.
It is precisely this term which determines the
nonzero value of the Casimir entropy at $T=0$K in the case of the Drude 
model \cite{6}. By repeating the derivation of Ref.~\cite{6} we conclude 
that the approach of Ref.~\cite{1} also violates the Nernst theorem,
as it leads to the following value of the entropy at $T=0$K:
\begin{equation}
S(0)=\frac{k_B}{4\pi}\int_{0}^{\infty}qdq\ln\frac{1-\left(
\frac{q-\delta_1}{q+\delta_1}\right)^2{\rm e}^{-2qd}}{1-\left(
\frac{q-\delta_2}{q+\delta_2}\right)^2{\rm e}^{-2qd}}<0.
\label{eq4}
\end{equation}
\noindent
Here, $\delta_1=(q^2+\tilde{\lambda}^{-2}+\omega_p^2/c^2)^{1/2}$,
$\delta_2=(q^2+\tilde{\lambda}^{-2})^{1/2}$, 
$\omega_p$ is the plasma frequency,
$\tilde{\lambda}=[\epsilon_0E_F/(3e^2c_t)]^{1/2}$
is the screening length at low temperature in the
Thomas-Fermi approximation \cite{8}, and $E_F$ is
the Fermi energy. 
Note that in the derivation of
Eq.~(\ref{eq4}) all Matsubara terms with both polarizations 
were taken into account so that the
TE modes with $n\geq 0$ have been modified, as prescribed by Eq.~(\ref{eq2})
suggested in Ref.~\cite{1}.

To conclude, we have compared the approach of Ref.~\cite{1} with the 
experimental data of two recent
precision experiments and found it to be inconsistent 
with these data at a high confidence level. 
This approach also violates the Nernst
heat theorem for a perfect crystal lattice, and is thus inconsistent with
thermodynamics.

R.S.D.~acknowledges NSF support through Grants No.~CCF-0508239
and PHY-0701636.
E.F. was supported in part by DOE under Grant No.~DE-76ER071428.
U.M., G.L.K. and V.M.M. were 
 supported by the NSF Grant No.~PHY0653657 
(computations of the Casimir pressure and force) and
DOE Grant No.~DE-FG02-04ER46131 (measurements of the Casimir force
between Au sphere and Si plate).
G.L.K. and V.M.M. were also partially supported by
Deutsche Forschungsgemeinschaft, Grant No.~436\,RUS\,113/789/0--4.

\begin{figure*}[h]
\vspace*{-7cm}
\centerline{
\includegraphics{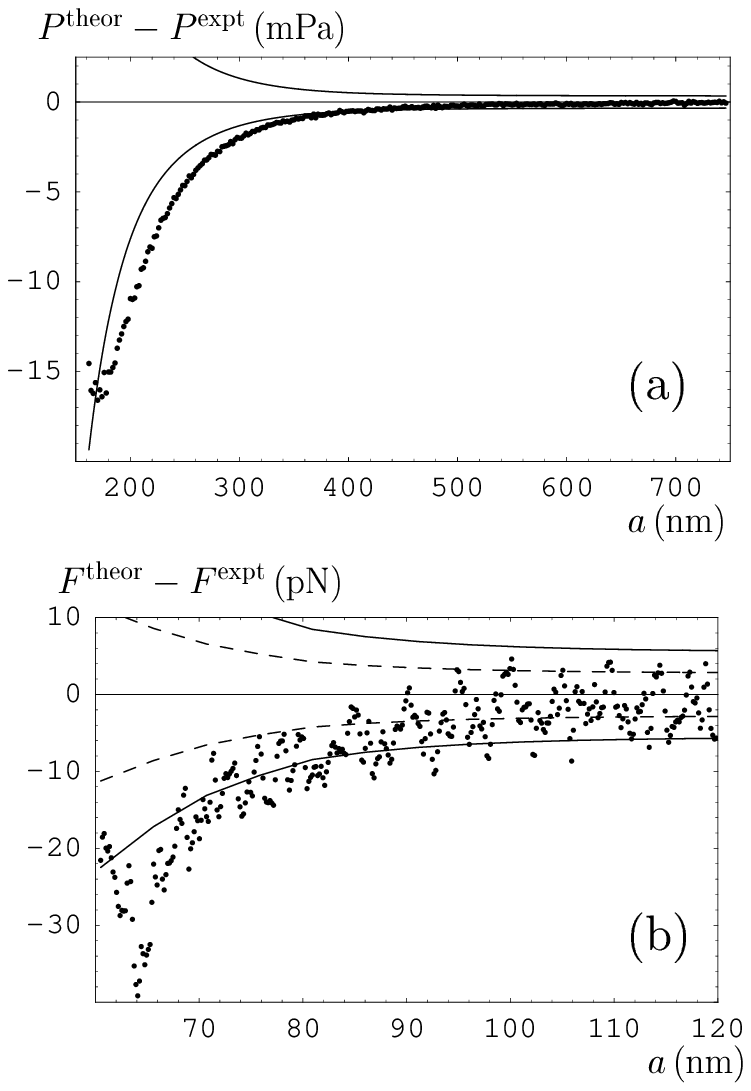}
}
\vspace*{-10cm}
\caption{
Differences between theoretical (using the approach of Ref.~\cite{1})
and experimental Casimir (a) pressures and (b) forces 
versus separation. The experimental data are taken from (a) Ref.~\cite{4}
and (b) Ref.~\cite{5}.
Solid and dashed lines indicate the borders of 95\% and 70\% confidence
intervals, respectively.
}
\end{figure*}
\end{document}